\documentclass[11pt]{article}
\usepackage[letterpaper,textwidth=6.5in,textheight=9in,
            centering,ignorehead,nomarginpar]{geometry}
\usepackage{color}
\usepackage{array}
\usepackage{natbib}
\bibpunct{(}{)}{;}{a}{,}{,}
\usepackage{graphicx}
\usepackage{bbm}
\usepackage{amsmath,amsthm,amssymb,amsfonts,latexsym}
\usepackage{booktabs}
\usepackage{icomma}
\usepackage{afterpage}
\usepackage[labelfont={small,sc},textfont={small,it},
            margin=15pt,skip=10pt,position=bottom]{caption}
\usepackage[labelfont={scriptsize,normalfont},textfont={scriptsize,it},
            margin=20pt,skip=5pt,position=bottom,
            labelformat=simple]{subcaption}
\usepackage{algorithmic}
\usepackage{algorithm}

\numberwithin{condition}{section}
\numberwithin{assumption}{section}
\numberwithin{remark}{section}
\numberwithin{equation}{section}
\numberwithin{lemma}{section}
\numberwithin{definition}{section}
\numberwithin{theorem}{section}
\numberwithin{proposition}{section}
\numberwithin{table}{section}
\numberwithin{figure}{section}
\numberwithin{theorem}{section}
\numberwithin{corollary}{section}
\numberwithin{property}{section}
\numberwithin{algorithm}{section}

\newcommand{\EQ}{\begin{equation}}
\newcommand{\EN}{\end{equation}}
\newcommand{\EQS}{\begin{equation*}}
\newcommand{\ENS}{\end{equation*}}

\def\n1{n}

\newsavebox{\savepar}

\numberwithin{equation}{section}
\numberwithin{table}{section}
\numberwithin{figure}{section}

\begin{document}
\title{
     Are target date funds dinosaurs? \\
     Failure to adapt can lead to extinction.
}

\author{Peter A. Forsyth\thanks{David R. Cheriton School of Computer Science,
        University of Waterloo, Waterloo ON, Canada N2L 3G1,
        \texttt{paforsyt@uwaterloo.ca}, +1 519 888 4567 ext.\ 34415.}
 \and
        Yuying Li\thanks{David R. Cheriton School of Computer Science,
        University of Waterloo, Waterloo ON, Canada N2L 3G1,
        \texttt{yuying@uwaterloo.ca}, +1 519 888 4567 ext.\ 7825.}
 \and
        Kenneth R. Vetzal\thanks{School of Accounting and Finance,
        University of Waterloo, Waterloo ON, Canada N2L 3G1,
        \texttt{kvetzal@uwaterloo.ca}, +1 519 888 4567 ext.\ 36518.}
}

\maketitle
\begin{abstract}
\vspace{10pt}
Investors in Target Date Funds are automatically switched
from high risk to low risk assets as their retirements approach.
Such funds have become very popular, but our analysis brings into 
question the rationale for them.  Based on both a model with 
parameters fitted to historical returns and on 
bootstrap resampling, we find that adaptive investment strategies 
significantly outperform typical Target Date Fund strategies.  This 
suggests that the vast majority of Target Date Funds are serving 
investors poorly.

\end{abstract}

\section{The pension problem}
Conventional defined benefit (DB) plans are becoming a thing of the 
past.  Most organizations do not want to take on the risk of providing 
a DB plan.  More employees are participating in defined contribution (DC) 
plans, and the trend is likely to continue.

In a typical DC plan, the employee contributes a fraction of his/her 
salary into a tax-advantaged savings account.  The employer may also 
contribute to the DC account.  In some cases, the employer manages the 
DC plan, in the sense that the employee picks from a list of approved 
investment vehicles, usually bond and stock mutual funds.
Upon retirement, the employee has to decide what to do with the 
accumulated amount in the portfolio.  Typical options include buying 
an annuity or continuing to manage the portfolio to generate a stream of 
income.  There is, of course, no guarantee of the level of income that 
will be produced in a DC plan.

It would not be unusual for a DC plan member to accumulate for thirty 
years (possibly with different employers), and then to decumulate for 
another twenty years.  This implies a fifty year investment cycle, making 
DC plan holders truly long term investors.

\section{Glide paths and constant proportions}
Before getting into some technical details, let's consider
two common investment strategies, and we will examine
how a DC investor would have fared during the 30 year
period from $1985-2015$.  We will assume that the 
investor had two possible assets in her DC fund: a short term
bond index  fund and a market capitalization weighted stock
index fund.  The investor had $\$10,000$ in $1985$ dollars to
start with, and contributed $\$10,000$ (in $1985$ dollars) each year
to the DC fund.

The simplest strategy is based on rebalancing
to a constant proportion stock-bond mix.  A typical
weight would be $60\%$ stocks and $40\%$ bonds.
This rebalancing to a constant mix was recommended by 
\citet{graham:2003} for defensive investors.

However, in order to avoid sudden drops in a DC portfolio
just before a retirement date, it is often suggested
that investors should use a {\em glide path} strategy.
In this case, we start off with a high allocation to
stocks, and then decrease the stock fraction as time
goes on.   We will consider a strategy where
the initial mix in 1985 is $80\%$ stocks and $20\%$ bonds
adjusting linearly over time to $20\%$ stocks and $80\%$ bonds
in $2015$.

We will compare these strategies with an {\em optimal adaptive}
strategy.  We will describe how we come up with this
strategy in Section \ref{adaptive_section}.  For now,
we'll just note that this adaptive strategy 
depends depends only on the total real wealth accumulated so far,
and on the time remaining before retirement.

\begin{figure}[tb]
\centerline{%
  \begin{subfigure}[b]{.4\linewidth}
    \centering
    \includegraphics[width=\linewidth]{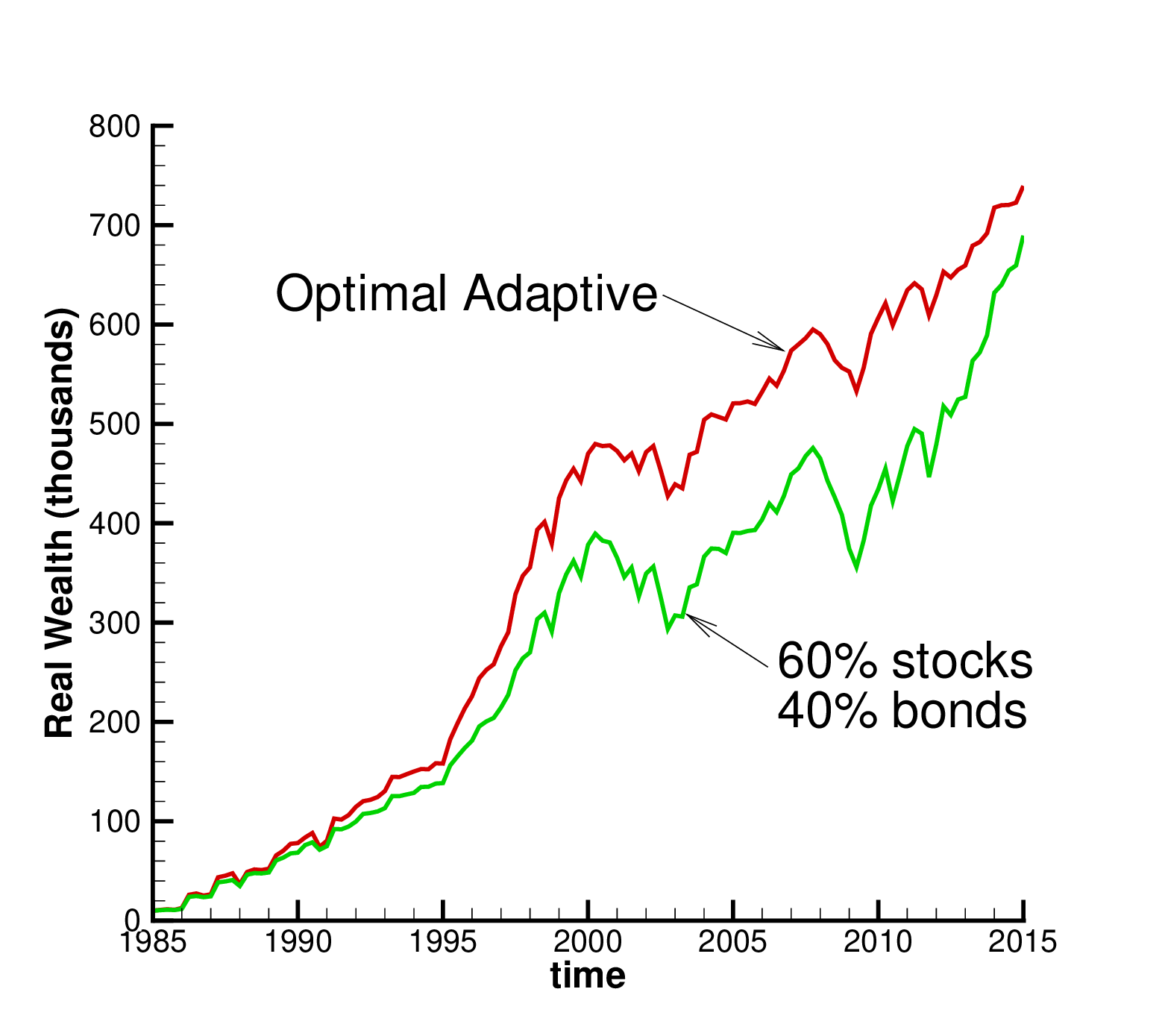}
    \caption{Comparison of optimal adaptive vs. constant proportion.
                \label{teaser_constant}}
  \end{subfigure}
  \begin{subfigure}[b]{.4\linewidth}
    \centering
    \includegraphics[width=\linewidth]{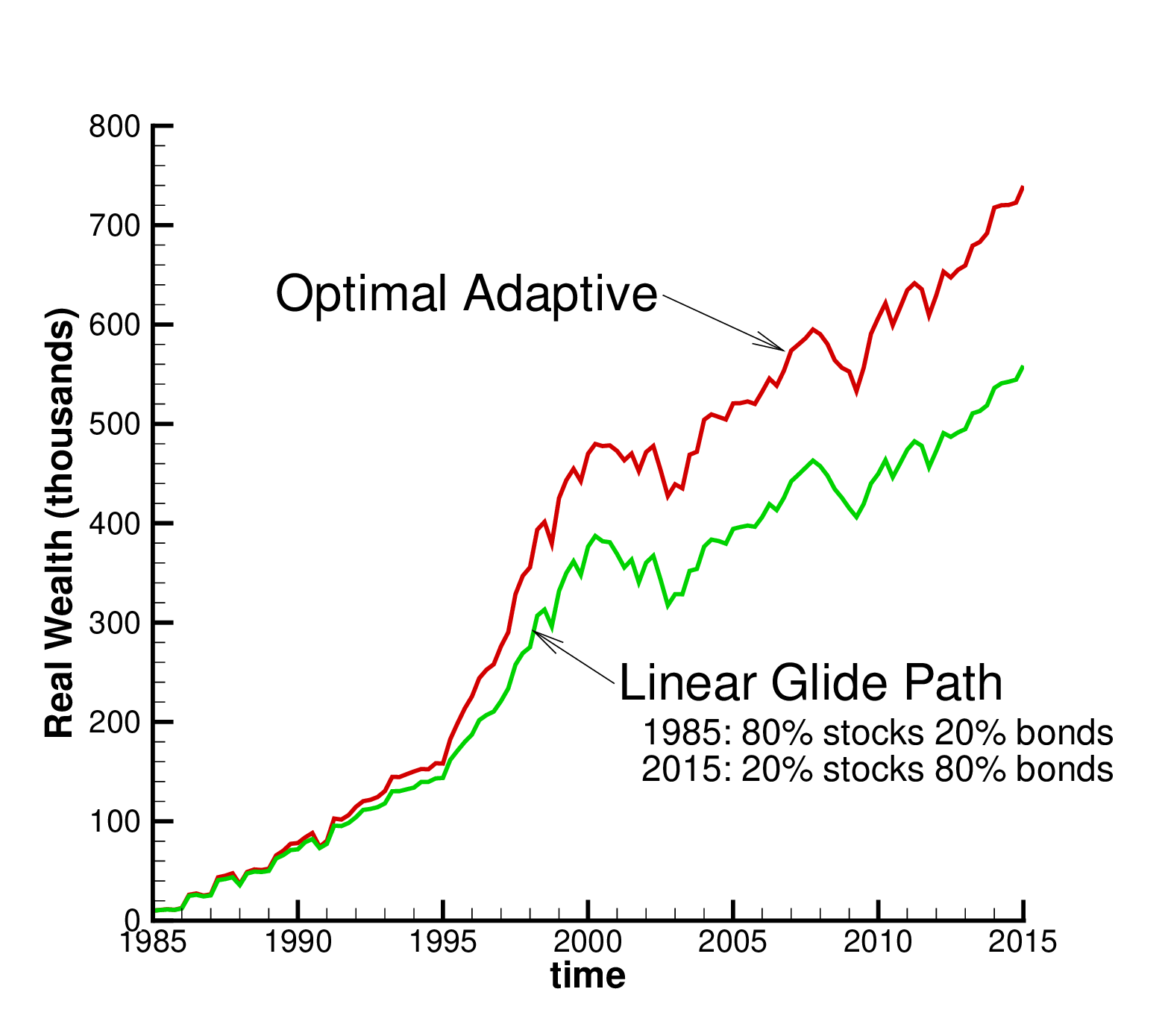}
    \caption{Comparison of optimal adaptive vs. linear glide path.
             \label{teaser_glide}}
  \end{subfigure}
  }
\caption{
Historical data: real (inflation adjusted) US total stock and short term bond returns, 1985:1 to 2015:12.
Linear glide path: 80\% stocks, 20\% bonds in 1985, dropping linearly over
time to 20\% stocks, 80\% bonds in 2015.  
Investor starts with  $\$ 10,000$ in $1985$ dollars, and
invests $\$ 10,000$ in $1985$ dollars each year. 
Optimal adaptive strategy described in
Section \ref{adaptive_section}.
Yearly rebalancing for all strategies.
\label{teaser}
}
\end{figure}

Figure \ref{teaser_constant} compares the adaptive strategy with
a constant proportion strategy.  We can see that
both strategies ended up with roughly the same total
real wealth, but the $60-40$ strategy had a very rough
ride during the dot-com bubble (2002) and the financial crisis (2008).

Alternatively, Figure \ref{teaser_glide} compares the adaptive
strategy with the linear glide path. 
As advertised, the glide path strategy is much smoother than 
the $60-40$ strategy, but it ends up with considerably smaller 
(i.e. about 30\% lower) real wealth in 2015 compared to the optimal adaptive strategy.

The optimal adaptive strategy appears to offer some advantages over 
both the constant proportion and the glide path strategies.
Perhaps you are intrigued.  How
does this strategy work?  We'll see in the next
Section.

\section{Three possible solutions}
Many studies have shown that individual investors generally do a poor 
job of investing.  They tend to buy at market peaks, sell at bottoms,
and are not well-diversified \citep[see, e.g.\@][]{barber2013}.
Target Date Funds (TDFs) (also known as Lifecycle Funds) are  an
attempt by the investment industry to provide a solution for retail clients, 
specifically those enrolled in a DC plan.\footnote{According to 
Morningstar, there was over USD 750 billion invested in TDFs in the US
at the end of 2015.} The most basic TDF has only two possible investments: 
a bond index and an equity index.  Given a specified target date (which 
would be the anticipated retirement date of the plan member), we consider 
here three possible methods to specify the bond/stock allocation in the
DC investment portfolio:  deterministic glide path,
constant proportion, and an adaptive strategy.

\subsection{Deterministic glide path}
In this case, the  allocation of stocks and bonds is determined
by a \emph{glide path}. This is currently a  popular method
used by many TDFs.  Denoting time by $t$, a simple example 
of a glide path is 
\[
\text{ Fraction invested in equities} = p(t) =  
  \frac{110 - \text{ your age at $t$}}{100}.
\]
The logic behind this idea is that you should take on more risk when you
are young (with many years to retirement) and then take on less risk
when you are older, with less time to recover from market shocks.  This
seems quite sensible.  The investment portfolio is typically rebalanced
at quarterly or yearly intervals, so that the equity fraction is
reset back to the glide path value $p(t)$.
This idea is so attractive that TDFs are \emph{Qualified Default 
Investment Alternatives} (QDIAs) in the US.  If an employee has enrolled 
in an employer-managed DC plan, the assets may be placed in a QDIA as a 
default option, in the absence of any instructions from the employee.

It is important to note that the glide path $p(t)$ in the age-based 
example above is only a function of time $t$.  We call this type of strategy 
a \emph{deterministic} glide path, i.e.\ this strategy
does not adapt to market conditions or the investment goals of the
DC plan member.

\subsection{Constant proportion}
A much simpler method is a \emph{constant proportion} policy, which 
is also a common asset allocation method. In this strategy, we
rebalance to a constant equity fraction $p_{const}$ at
all rebalancing times.  Of course, a constant proportion allocation is 
a special case of a glide path, where $p(t) = p_{const}$.

\subsection{Adaptive strategies}\label{adaptive_section}
In deterministic glide paths, rebalancing strategies are only
a function of time.  Let's consider a strategy which allows the 
fraction invested in the risky asset to be a function of both
time $t$ and accumulated wealth in the DC portfolio at $t$, denoted by 
$W_t$.  Then $p = p(W_t,\:t)$, so that this an \emph{adaptive} strategy. 

We consider a target-based strategy, where we choose
$p(W_t,t)$ to minimize
\begin{equation}
E \left[ \left(W_T - W^*\right)^2 \right],
\label{objective_1}
\end{equation}
where $W_T$ denotes terminal wealth at time $T$, $W^*$ is target final
wealth, and $E[\cdot]$ indicates expected (or mean) value.
In other words, we seek the asset allocation strategy which minimizes
the expected quadratic shortfall with respect to the target
wealth $W^*$.\footnote{We assume that the initial
wealth $W_0 < W^*$.  \citet{Dang2015a} show that
the optimal strategy has $W_t \leq W^*$,
so that $W_T \leq W^*$.  This means that only shortfall (not excess) is
penalized in (\ref{objective_1}).}

\section{Comparing the three solutions}
To provide a realistic comparison, we consider a plausible investment scenario
and evaluate the three strategies under both (i)~a parametric model 
which captures the broad statistical properties of
the historical market, and (ii)~bootstrap resamples of the historical market.

\subsection{Investment scenario}
We consider the prototypical DC investor example shown in
Table~\ref{base_case}.  We assume that the investor makes an
initial investment in the portfolio of \$10,000 at time zero 
(i.e.\ $W_0 = \$10,000$), and
that she contributes \$10,000 per year (measured in real terms,
i.e.\ inflation-adjusted) to the DC fund.  The investment horizon
considered is $T=30$ years, with the last investment of \$10,000
(real) being made at $t=29$ years.

\begin{table}[tb]
  \begin{center}
    \begin{tabular}{ll} \toprule
      Investment horizon & 30 years \\
      Synthetic market parameters & Historical data (1926:1-2015:12)\\
      Initial investment $W_0$ & \$10,000 \\
      Real investment each year & \$10,000 \\
      Rebalancing interval & 1 year \\ \bottomrule
    \end{tabular}
    \caption{Long term investment scenario. After the initial investment,
      cash is injected and rebalancing occurs at $t = 1,\ldots, 29$ 
      years.\label{base_case}}
  \end{center}
\end{table}

\subsection{Estimating a parametric model for real equity returns}
We construct two real indexes: a real total return equity
index and a real short term bond index.  Our data was obtained from
the Center for Research in Security Prices (CRSP) through
Wharton Research Data Services.\footnote{More
specifically, results presented here were calculated based
on data from Historical Indexes, \copyright 2015 Center for Research
in Security Prices (CRSP), The University of Chicago Booth School of
Business.  Wharton Research Data Services was used in preparing this article.
This service and the data available thereon constitute valuable intellectual
property and trade secrets of WRDS and/or its third-party suppliers.} We
use the CRSP value-weighted total return index (``vwretd''), which includes
all distributions for all domestic stocks trading on major US exchanges.
We also use the 30-day Treasury bill return index from CRSP.  Both this index
and the equity index are in nominal terms, so we adjust them for
inflation by using the US CPI index (also supplied by CRSP).  
We use real indexes since long term retirement saving should be trying
to achieve real (not nominal) wealth goals.

\begin{figure}[tb]
\centerline{%
  \begin{subfigure}[b]{.55\linewidth}
    \centering
    \includegraphics[width=\linewidth]{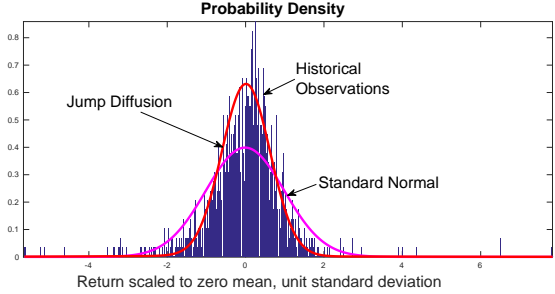}
    \caption{Probability densities.\label{density_1}}
  \end{subfigure}
  \begin{subfigure}[b]{.55\linewidth}
    \centering
    \includegraphics[width=\linewidth]{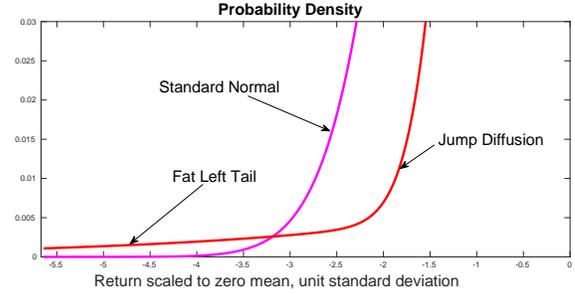}
    \caption{Zoom of Figure \ref{density_1}, showing the fat left tail.
             \label{density_2}}
  \end{subfigure}
  }
\caption{Probability density of monthly returns for real CRSP VWD index.
Monthly data, 1926:1 - 2015:12, scaled to unit standard deviation and zero mean.
Standard normal density and fitted jump diffusion model also shown.
\label{monthly_hist_1}
}
\end{figure}

Figure~\ref{density_1} shows a histogram of the monthly
returns from the real total return equity index, scaled to
unit standard deviation and zero mean.  We superimpose a standard
normal (Gaussian) density onto this histogram.  The plot shows that
the empirical data has a higher peak and fatter tails than a normal
distribution, consistent with previous empirical findings for virtually
all financial time series.

The \emph{fat left tails} of the historical density function
can be attributed to large downward equity price movements which are
not well modelled assuming normally distributed  returns.
From a long term investment perspective, it is advisable to
take into account these sudden downward price movements.

We fit the data over the entire historical period using a jump
diffusion model \citep{Kou2004}.  This provides a more accurate fit to the
data, as illustrated in Figure~\ref{density_1}.  To show the fat left
tail of the jump diffusion model, we have zoomed in on a portion
of the fitted distributions in Figure~\ref{density_2}.

\subsection{Comparisons under the estimated model}
We now compare the constant proportion, (optimal) deterministic glide path 
and optimal adaptive strategies under the estimated jump diffusion model with 
parameters estimated from the entire 1926:1 - 2015:12 data set (subsequently 
referred to as the \emph{synthetic market}).  Note that the jump diffusion
model is applied only to the equity returns.  Bond returns
are simply determined from the sample average monthly change in the real 
bond index.  We will use this synthetic market to determine optimal strategies
and carry out Monte Carlo simulations.

In the following comparison, we consider a $60{-}40$ equity-bond split 
as the constant proportion strategy.  In other words, at each annual
rebalancing date we rebalance so that 60\% of the portfolio is invested
in equities and 40\% in bonds.  This is a special case
of a glide path strategy, with $p(t) = .60$ for all times $t$.
The equity fractions at rebalancing times for the other two strategies
are determined as follows:

\begin{itemize}
\item Deterministic glide path strategy:   we calculate the equity
fraction at each rebalancing date such that the standard deviation of
terminal wealth $std[W_T]$ is as small as possible under the 
restriction that the expected value of terminal wealth $E[W_T]$ matches
the corresponding expected value for the constant proportion strategy.
\item Adaptive strategy:  we calculate the equity fraction at each
rebalancing date such that the mean quadratic target error
$E\left[\left(W_T - W^*\right)^2\right]$ is as small as possible with
the target $W^*$ set so that the expected value of terminal wealth 
$E[W_T]$ is the same as for the constant proportion strategy.
\end{itemize}
We determine the optimal rebalancing fractions by using a computational
optimization method (glide path) and solving a Hamilton Jacobi Bellman
equation (adaptive strategy).  In each case, we constrained the equity 
fraction $p$ so that $0 \leq p \leq 1$ (no shorting and no leverage).

\begin{figure}[tb]
\centerline{%
  \begin{subfigure}[b]{.4\linewidth}
    \centering
    \includegraphics[width=\linewidth]{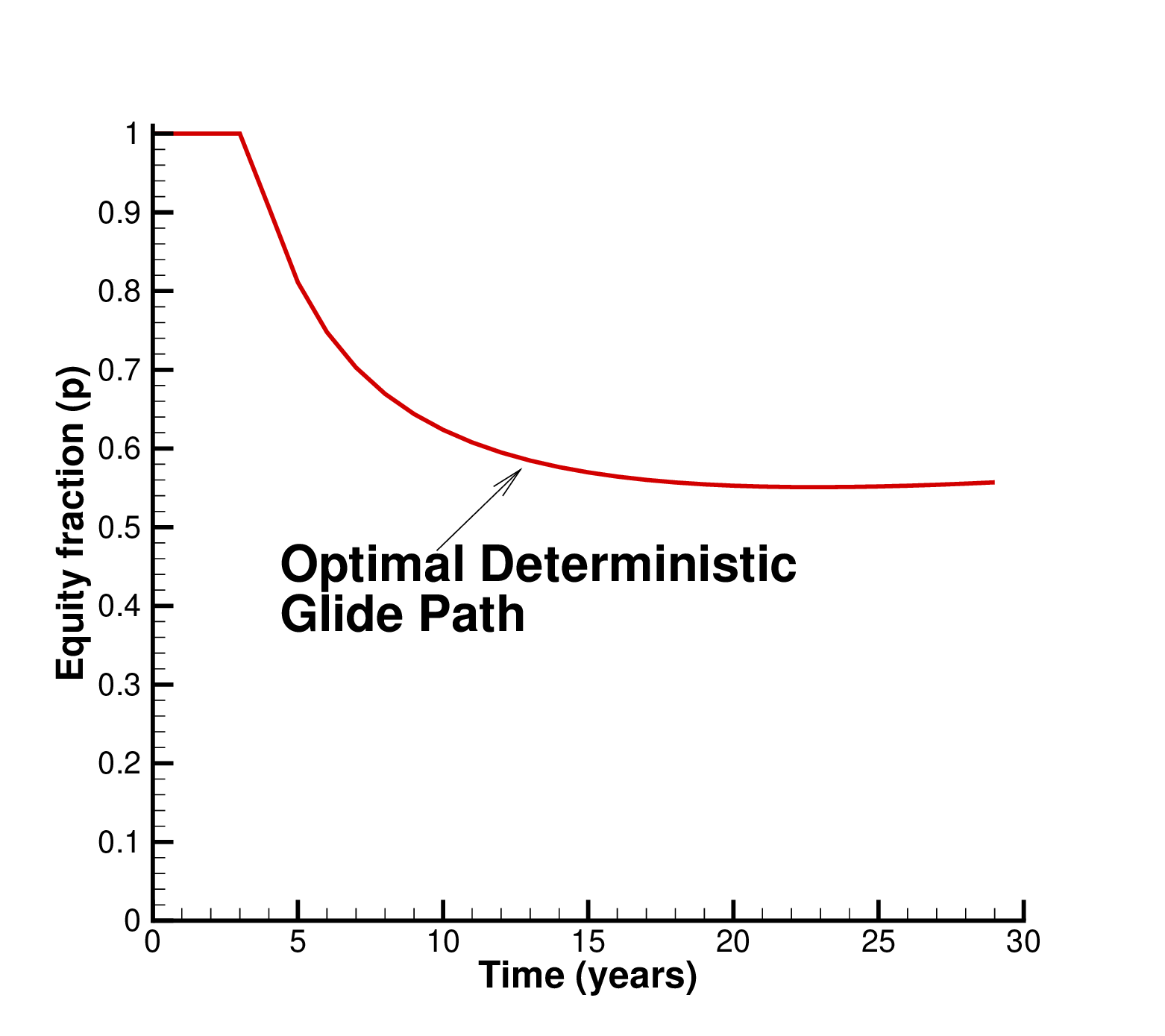}
    \caption{Optimal deterministic glide path.\label{deterministic_path_1}}
  \end{subfigure}
  \begin{subfigure}[b]{.4\linewidth}
    \centering
    \includegraphics[width=\linewidth]{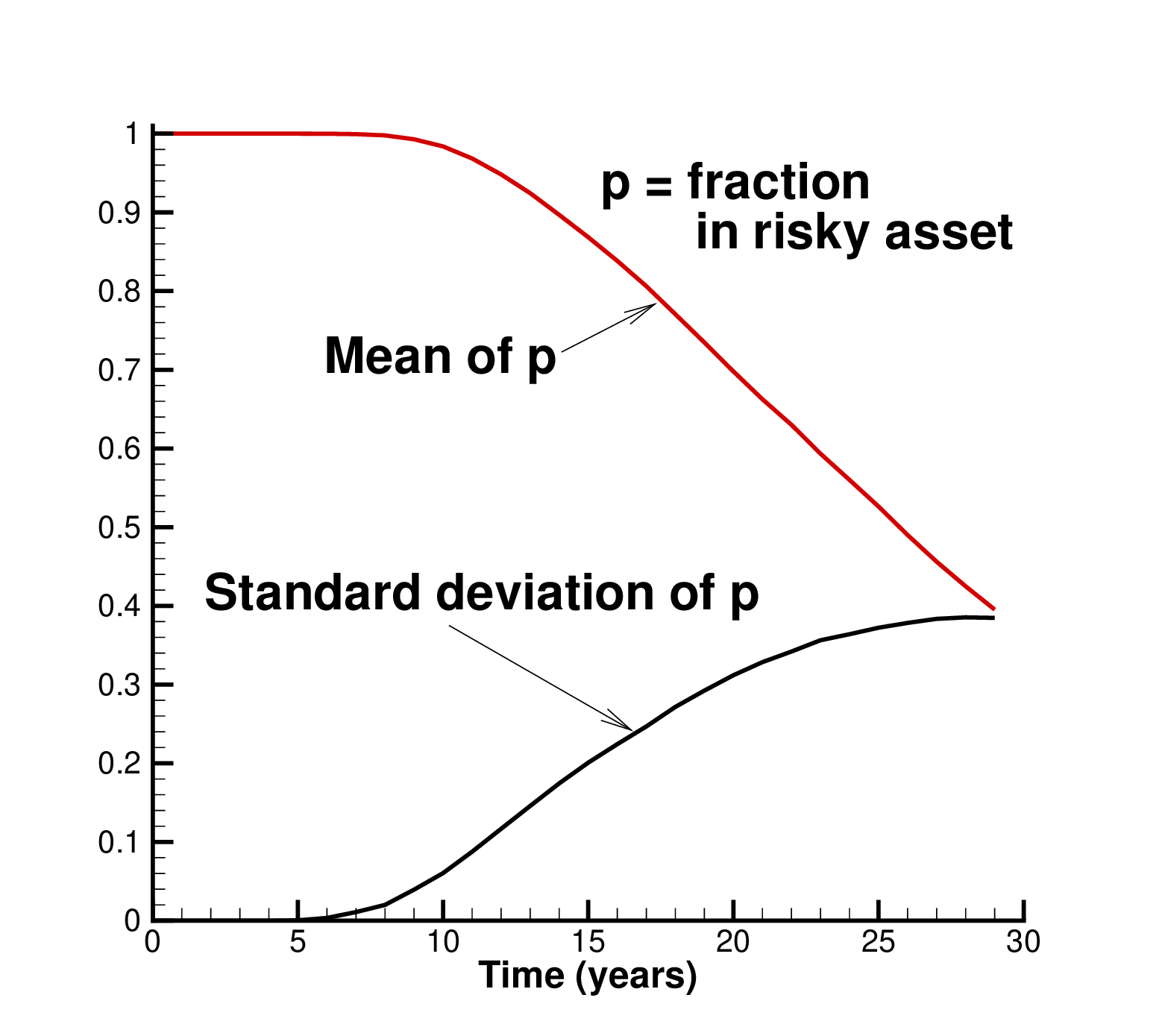}
    \caption{Optimal adaptive path.
             \label{adaptive_path}}
  \end{subfigure}
  }
\caption{
Optimal deterministic and adaptive paths.  For the adaptive case,
we show the expected value of the fraction invested in the
risky asset $(p)$ and its standard deviation.  The adaptive
case is shown based on Monte Carlo simulations.
\label{optimal_paths}
}
\end{figure}

Figure \ref{deterministic_path_1} shows the optimal deterministic glide path.
Figure \ref{adaptive_path} displays the mean fraction invested in
equities for the adaptive strategy, and its standard deviation.\footnote{Since
the adaptive strategy depends on the accumulated wealth so far, it is
non-deterministic.  As a result, we plot both the mean and
the standard deviation of the equity fraction.}
On average, the adaptive strategy maintains a high allocation
to stocks for longer than the deterministic strategy, but
de-risks faster as we approach retirement.
The constraint $p \leq 1$ is clearly active for both strategies at 
early times.

Table \ref{base_results} compares the results for the constant 
proportion $(p=.60$), optimal deterministic glide path, and adaptive
strategies.  All three strategies by design deliver the same 
expected value.  The optimal deterministic
standard deviation is about $0.97$ of the constant
proportion strategy, a very small
improvement. The standard deviations of the glide path and constant
proportion are more than twice as large as that of the adaptive strategy.
Table~\ref{base_results} also shows some information about shortfall
probabilities.  For example, the probability of achieving
a final real wealth less than \$650,000 
is about 43\% for both the constant proportion the and optimal deterministic
strategies. For the adaptive strategy, the probability of achieving
a final wealth less than \$650,000 is only 22\%.

Based on these synthetic market results for this case with
regular contributions, it is possible to come up with
a deterministic glide path which beats a constant
proportion strategy, but not by very much. The adaptive strategy, 
on the other hand, significantly outperforms both
the optimal deterministic glide path and the constant proportion
strategies. We have repeated these tests with many different synthetic 
market parameters.  As long as the investment horizon is longer than 20 years,
the differences in performance between the optimal deterministic
glide path and the constant proportion strategy with equivalent terminal
wealth are very small, and each of these are clearly dominated
by the adaptive strategy.

\begin{table}[tb]
\begin{center}
{\small
\begin{tabular}{lccccc} \toprule
 & & & \multicolumn{3}{c}{Probability of Shortfall} 
 \\ \cmidrule{4-6}
Strategy & $E[W_T]$ & $std[W_T]$ & $W_T < \$500,000$ & $W_T < \$650,000$
 & $W_T < \$800,000$ \\ \midrule
Constant proportion & \$824,000 & \$512,000 & .23 & .44 & .60 \\
Deterministic glide path & \$824,000 & \$503,000 & .23 & .43 & .60 \\
Optimal adaptive & \$824,000 & \$242,000 & .15 & .22 & .31 \\ \bottomrule 
\end{tabular}
}
\caption{Scenario data in Table~\ref{base_case}.  $W_T$ is accumulated
real portfolio wealth at $T=30$ years.  Synthetic market results 
for a constant proportion strategy with $p=.60$, an optimal deterministic
glide path strategy, and an optimal adaptive strategy.
\label{base_results}
}
\end{center}
\end{table}

\subsection{Comparisons under  bootstrap resampling}
The synthetic market results are based on fitting the
historical returns to a jump diffusion model.  This model
assumes that monthly equity returns are statistically independent, which is
debatable.  In order to get around
artifacts introduced by our modelling assumptions,
we test the three strategies using a \emph{bootstrap resampling} method.
This can be viewed as a more realistic test, in the sense
that we can observe how the strategies would have performed
on actual historical data.

Our investment horizon is $T = 30$ years.
Each bootstrap path is determined by dividing $T$
into $k$ blocks of size $b$ years, so that $T = kb$.
We then select $k$ blocks at random (with replacement) from
the historical data set.  Each block starts at a random month.
We then concatenate these blocks to form a single path.
We repeat this procedure $10,000$ times and generate statistics
based on this resampling method.

The idea here is that if the real data shows some serial dependence,
then this will show up in the bootstrap resampling.  Based on some 
econometric criteria, we use a blocksize of $2.0$ years.  We have 
experimented with blocksizes ranging from $0.5-10.0$ years, and the 
results are qualitatively similar.  Note that the rebalancing
strategies continue to be determined from the synthetic market as
described above.  We emphasize that the resampled paths are
applied to both the equity index and the bond index. In addition
to serial dependence, we are also introducing random variations 
in the bond component through the resampling.

The bootstrap results are shown in Table~\ref{bootstrap_table_cap_wt}.
Of course, now that we use the actual historical returns (with a random 
starting point), the mean terminal wealth for the adaptive strategy 
is no longer equal to the mean terminal wealth for the constant 
proportion strategy.  This reflects the fact that our strategies were 
computed based on the jump diffusion model with 
constant interest rates, which is obviously an imperfect
representation of reality.
Nevertheless, as for the synthetic market tests,
the adaptive strategy significantly dominates
both the constant proportion and the deterministic glide path strategies.

We emphasize that the adaptive strategy was based only on the
the long term average parameters obtained by fitting a jump
diffusion model to the entire historical data set.  This same strategy
was then used for all the simulated bootstrap resampled
paths. No tuning of the strategy for any particular path
was used.

\begin{table}[tb]
\begin{center}
{\small
\begin{tabular}{lccccc} \toprule
 & & & \multicolumn{3}{c}{Probability of Shortfall} 
 \\ \cmidrule{4-6}
Strategy & $E[W_T]$ & $std[W_T]$ & $W_T < \$500,000$ & $W_T < \$650,000$
 & $W_T < \$800,000$ \\ \midrule
Constant proportion & \$784,000 & \$390,000 & .23 & .44 & .61 \\
Deterministic glide path & \$784,000 & \$382,000 & .22 & .43 & .62 \\
Optimal adaptive & \$814,000 & \$229,000 & .14 & .21 & .33 \\ \bottomrule 
\end{tabular}
}
\caption{Example in Table \ref{base_case}.  $W_T$ is accumulated
real portfolio wealth at $T=30$ years.  Bootstrap resampling results
based on historical data from Jan.\ 1926 to Dec.\ 2015 for a constant
proportion strategy with $p=.60$, an optimal deterministic glide path strategy,
and an optimal adaptive strategy.  10,000 bootstrap resamples were used,
with a blocksize of 2 years.
\label{bootstrap_table_cap_wt}
}
\end{center}
\end{table}

\subsection{More on the optimal adaptive strategy}
The target $W^*$ for the adaptive strategy was selected so that expected
terminal wealth in the synthetic market matched that achieved by the
constant proportion strategy.  In practice, how should we pick $W^*$?
A reasonable approach is to enforce the constraint
\begin{equation}
  E[W_T] = W_{\text{\emph{goal}}}
         = \text{ investment goal}~.
       \label{constraint_1}
\end{equation}
The investment goal in the case of retirement saving would be the
amount required to fund a reasonable replacement level of income for
a retiree.  Of course in general, $W_{\text{\emph{goal}}} < W^*$, 
i.e.\ in order to have $W_{\text{\emph{goal}}}$ wealth on average, 
we have to aim at a higher target.

It is also interesting to note that the adaptive strategy
turns out to be dynamically
mean variance optimal \citep{li-ng:2000,Dang2015a}.
This means that for a specified mean terminal wealth $E[W_T]$,
no other strategy has smaller variance.  In addition, the
adaptive strategy offers the opportunity in some cases to withdraw 
cash without compromising the probability of reaching the target
\citep{Dang2015a,forsyth2016abc}.

\section{Conclusion: Do target date funds need to adapt?}
Of course, we have looked at only one possible adaptive
strategy.  There are many other possibilities.  We argue
that the adaptive strategy we have considered is especially
appealing because it simultaneously minimizes two measures of risk:
quadratic shortfall and variance (standard deviation).

However, our main point here is that restricting attention
to deterministic glide paths is suboptimal.  Investors can do
a lot better by considering adaptive strategies.  It is worthwhile
to note that the vast majority of target date funds use
deterministic strategies.  Do target date funds need to adapt?  The 
answer is clear: you need to adapt to your target.


\end{document}